\documentclass[12pt,a4paper]{article}
\usepackage{amsmath,latexsym}
\begin{document}

\title{ 
{\large Review of the FoPL paper \cite{Evans 1}}\\
The Evans Lemma of Differential Geometry}

\author{Gerhard W.~Bruhn\\
Darmstadt University of Technology\\ 
D 64289 Darmstadt, Germany\\
{\small bruhn@mathematik.tu-darmstadt.de}}


\maketitle

\begin{abstract}

{\footnotesize
The Evans Lemma is a basic tool for Evans GCUFT or ECE Theory \cite{Evans 2}. Evans has given
two proofs of his Lemma. The first proof in \cite{Evans 1} is shown to be invalid due to 
dubious use of the covariant derivative $D^\mu$. A second proof in
\cite[Sec.J.3]{Evans 2} is wrong due to a logical error.}
\end{abstract}

\vspace*{1cm}


\newpage

\section{{\sc M.W. Evans}' first proof of his Lemma}

The Evans Lemma is the assertion of proportionality of the matrices
$(\Box q^a_\mu)$ and $(q^a_\mu)$ with a proportionality factor $R$:
$$
(\Box q^a_\mu) = R (q^a_\mu) .
$$\\

{\em Quotation from} \cite[p.432+8]{Evans 1}\footnote{The page numbers of the web copy
mentioned in \cite{Evans 1} start with 1 instead of 433 (= 432+1).}\\

{\footnotesize
The Evans lemma is a direct consequence of the tetrad postulate.
The proof of the lemma starts from covariant differentiation of the
postulate:\\

$\cite[(36)]{Evans 1}\footnote{Quotations from {\sc M.W. Evans}' contributions \cite{Evans 1}, 
\cite{Evans 2} and \cite{Evans 3} appear with equation labels [p,(nn)] in the left margin.}
\hspace*{2.5cm} 
D^\mu (\partial_\mu q^a_\lambda  + 
\omega^a_{\mu b}q^b_\lambda  -   
\Gamma^\nu_{\mu\lambda} q^a_\nu) = 0. $\\
 
Using the Leibnitz rule, we have\\

$\cite[(37)]{Evans 1}\hspace*{0.8cm} 
(D^\mu \partial_\mu )q^a_\lambda  +
\partial_\mu (D^\mu q^a_\lambda  ) + 
(D^\mu \omega^a_{\mu b})q^b_\lambda   
+
\omega^a_{\mu b}(D^\mu q^b_\lambda  )
-  (D^\mu  \Gamma^\nu_{\mu\lambda} )q^a_\nu  -   
\Gamma^\nu_{\mu\lambda}  (D^\mu q^a_\nu  ) = 0,$\\
 
and so\\

\cite[(38)]{Evans 1}\hspace*{2.8cm} 
$
(D^\mu \partial_\mu )q^a_\lambda  + 
(D^\mu \omega^a_{\mu b})q^b_\lambda  -  
(D^\mu  \Gamma^\nu_{\mu\lambda} )q^a_\nu  = 0,
$\\ 
 
because\\

\cite[(39)]{Evans 1}\hspace*{2.8cm} 
$
D^\mu q^a_\lambda  = 
D^\mu q^b_\lambda  = 
D^\mu q^a_\nu  = 0.
$\\
}
 
{\em End of Quotation}\\

Eq.\cite[(36)]{Evans 1} is formally correct, however, the decomposition in 
Eq.\cite[(37)]{Evans 1} yields {\em undefined}
expressions: What e.g. is the meaning of the terms
$D^\mu \omega^a_{\mu b}$ and $D^\mu \Gamma^\nu_{\mu\lambda}$?
Note that both $\omega^a_{\mu b}$ and $\Gamma^\nu_{\mu\lambda}$ are no tensors 
and so the covariant derivative $D^\mu$ 
is not applicable. Therefore we skip over the rest of [1].\\


\section{{\sc Evans}' second proof of his Lemma}

{\sc M.W. Evans} himself felt it necessary to give another proof in \cite[p.514]{Evans 2}, 
now avoiding the problem of undefined terms.\\

{\em Quotation from} \cite[p.514]{Evans 2}\\

{\footnotesize
{\bf J.3 The Evans Lemma}\\

The Evans Lemma is the direct result of the tetrad postulate of differential
geometry:\\
 
\cite[(J.27)]{Evans 2}
\hspace*{2.8cm}
$
D_\mu q^a_\lambda  =
\partial_\mu q^a_\lambda  + 
\omega^a_{\mu b}q^b_\lambda  -   
\Gamma^\nu_{\mu\lambda} q^a_\nu  = 0.
$\\ 
 
using the notation of the text. It follows from eqn. (J.27) that:\\
 
\cite[(J.28)]{Evans 2}
\hspace*{2.8cm}
$
D^\mu (D_\mu q^a_\lambda ) = 
\partial^\mu (D_\mu q^a_\lambda ) = 0, 
$\\
 
i.e.\\
 
\cite[(J.29)]{Evans 2}
\hspace*{2.8cm}
$
\partial^\mu  (\partial_\mu q^a_\lambda  + 
\omega^a_{\mu b}q^b_\lambda  -   
\Gamma^\nu_{\mu\lambda} q^a_\nu ) = 0, 
$\\
 
or\\
 
\cite[(J.30)]{Evans 2}
\hspace*{2.8cm}
$
\Box q^a_\lambda  = 
\partial^\mu (\Gamma^\nu_{\mu\lambda} q^a_\nu ) 
- 
\partial^\mu (\omega^a_{\mu b}q^b_\lambda ) . 
$\\

Define:\\

\cite[(J.31)]{Evans 2}
\hspace*{2.8cm}
$R q^a_\lambda  := 
\partial^\mu (\Gamma^\nu_{\mu\lambda} q^a_\nu ) 
-  
\partial^\mu (\omega^a_{\mu b}q^b_\lambda)
$\\
 
to obtain the Evans Lemma:\\
 
\cite[(J.32)]{Evans 2}
\hspace*{2.8cm}
$
{\bf \Box} q^a_\lambda  = R q^a_\lambda
$\\  
}

{\em End of Quotation}\\

{\bf As simple as wrong}: 
Eq.\cite[(J.31)]{Evans 2} represents a set of 16 equations
each of which for one fixed pair of indices $(a,\mu) (a,\mu = 0,1,2,3)$. 
Each equation is a condition to be fulfilled by the quantity $R$. These 16 conditions
for $R$ do {\em not agree} in general.\\

Thus, the author Evans, when giving the "definition" \cite[(J.31)]{Evans 2},
ignored the possible incompatibility of the {\em sixteen} definitions of $R$ contained in his
"definition" of $R$ by Eq.\cite[(J.31)]{Evans 2}. Therefore this proof of the Evans Lemma 
in \cite[Sec.J.3]{Evans 2} is invalid.\\

{\bf Conclusion: There is no proof of the Evans Lemma, neither in the article \cite{Evans 1}
 nor in \cite[Sec.J.3]{Evans 2}.}

{\bf Additional remark} In his note \cite[p.2]{Evans 3} Evans gives a variation of this "proof". 
There he defines $R$ directly and applies his "Cartan Convention":\\
 
{\em Quotation from} \cite{Evans 3}\\

{\footnotesize
\cite[(9)]{Evans 3}
\hspace*{2.8cm}
$
R = 
q^\lambda_a \partial^\mu (\Gamma^\nu_\mu\lambda q^a_\nu 
- \omega^a_{\mu b}q^b_\lambda)
$\\

and use $<$the "Cartan Convention"$>$\\
 
\cite[(10)]{Evans 3}
\hspace*{2.8cm}
$
q^\lambda_a q^a_\lambda = 1
$\\
 
to find\\
 
\cite[(11)]{Evans 3}
\hspace*{2.8cm}
$
{\bf \Box} q^a_\lambda = R q^a_\lambda .
$\\
}

{\em End of Quotation}\\

i.e. from the correct Eq. [2,(J.30)] he {\em erroneously} concludes\\
 
$$
q^a_\lambda R = 
(q^a_\lambda q^\lambda_a) \partial^\mu (\Gamma^\nu_\mu\lambda q^a_\nu 
- \omega^a_{\mu b}q^b_\lambda)
= 1 \cdot {\bf \Box} q^a_\lambda
$$\\

We learn from this that one can "prove" every nonsense, if one has the suitable
error at hand, e.g. ignore the rules of tensor calculus on hidden indices.
(see also \cite[Evans' New Math in Full Action ...]{Bruhn})\\


\end{document}